# POTENTIAL USE OF IOT DISTANCE MEASUREMENT TOOL IN BOULE SPORTS

Wahidah Md Shah[1], M Azim. Adnan[1], Aslinda Hassan[1], Norharyati Harum[1] and Isredza Rahmi A. Hamid[2]

[1]Fakulti Teknologi Maklumat dan Komunikasi (FTMK), Universiti Teknikal Malaysia Melaka (UTeM).
[2]Fakulti Sains Komputer Dan Teknologi Maklumat, Universiti Tun Hussein Onn.

*ABSTRACT*

*In Petanque, each player aims to throw the boule closer to the jack. The closest boule to the jack among players will score the point. Currently, the distance of the boule to the jack is still measured using manual measurement tools such as measuring tape, string, and calipers. The manual measurement method is considered time-consuming and prone to inconsistent reading, which the ordinary referees and players conduct. A steady hand is required to hold the tape at two ends while squatting or kneeling. The technique of reading the measurement is also important to determine the accuracy of the length. This project aims to design and develop a prototype device that can measure the distance between jack and boule using a microcontroller and ultrasonic sensor technology. The device is expected to provide an instant measurement of the distance between the jack and the boule. The measurement data can be displayed on the mobile device to ease the user to view the result. This prototype device also counts the score points and determines the winner.*

## 1. INTRODUCTION

The Internet of Things (IoT) has emerged as a key enabler for smart technology that allows communication and data exchange among devices in a network, such as sensors, microcontrollers, and mobile devices. The emergence and adaptability of IoT in daily tasks will bring changes in completing a task. On the other hand, IoT also catches great attention among academic researchers who study the potential and performance of IoT in various fields, including health. [1], disabled [2], surveillance [3], patient monitoring [4] and even in sport [5]. For example, an IoT-based system can monitor patients' health and alert the medical staff if any abnormality happens. In disaster management, the authorities can be informed instantly when the volcano shows the potential to burst; thus, evacuation can be made promptly. A gate intelligence system that can alarm intruders' owners is an example of a surveillance system. Sports and IoT are also potential fields to explore. There are some related works in sports such as in [4], researchers propose monitoring athlete performance in cycling sport or even assisting competition in marathon event [6].

We are interested in exploring the possibility of automating the measurement of distance score types of sports such as petanque, long and high jump, and javelin throw. This paper presents Petanque sport as a case study. Petanque is a sport that is classified as a boules sport, in which the player will play the ball (boule) towards a target ball (jack). The game score point in Petanque sport is based on the nearest distance between the thrown boule and the targeted jack. For example, in team A with two players, both players' distances are 2 and 3 meters while the opponent's distance is 4 meters. Team A earned two points for both balls by the players.

There are several ways to determine which boule has the closest distance to the jack [7]. One of the techniques is called visual inspection. This technique is rarely used for real games since the distance is not accurately identified. The reading must be done by standing in the designated area to determine which two boules are closest to the jack. It is fast, and there is no risk of touching and moving the boule, but this is not precise. To ensure a fair game, proper measurement needs to be done.

Besides, there is a comparison technique where the distance between the boules and the jack is compared using a physical object whose length can be adjusted. Various objects can be used to make the comparison, and some examples are calipers, string, and folding rulers. Unfortunately, this technique is not practical for long distances. The last technique and the most common one is the measurement technique. A tape measure is the most common tool for measuring distances between the jack and the boules; after the measurements are collected, they will be compared to determine which boule is closer. The measurement technique is accurate if the measurement technique is correct. It required a steady hand to hold the tape measure with two different parts of the tape while squatting or kneeling, then reading the measurement to the accuracy of one millimeter while avoiding distortions. The risk of moving boule and jack is high for this technique, and it can lead to inconsistent results and be difficult for ordinary players.

Moreover, manual measurement is prone to human error. Therefore, this project explores the usage of IoT to facilitate the measuring technique and reduce human error. A prototype jack is designed to measure the distance between the jack and boule, with a mobile application displaying the score and winner.

The paper is organized as follows. Section 2 presents the related works of IoT in sports; Section 3 elaborates on the methodology used in developing the prototype. Section 4 analyses the results and discusses them accordingly. Section 5 concludes the paper with an outline of contributions and future work.

## 2. RELATED WORK

IoT in sports has been researched for several aims, including training, monitoring performance, assisting games, and analytic purposes. Different sports require different sensors, types of data to be collected, processing methods, etc. The idea of adapting IoT in sports for performance monitoring is discussed by [8]. They highlighted the possible sensors and gadgets that can be used, the critical information to be collected, the data processing, etc., which required future research on optimizing IoT for high potential.

Most IoT-based sports systems aim to monitor the athlete's performance, especially during training sessions. For example, for bowling games in [9], inertial measurement unit (IMU) sensors capture the bowler's motion while throwing. The motion data helps make coaching sessions easier by assessing the throwing performance. Similarly, in basketball games, IMUs also can be used to capture a player's shooting style. In [10] two IMUs are used; one is custom-built for wrist-worn prototypes, and another is embedded in the basketball. The sensors capture motion, while an additional camera matches the shooting style. Both data influence shooting quality, easing monitoring and coaching. While [11] also captures movement data for dumbbell players. It categorizes the movement of the dumbbell, whether standard or non-standard, facilitating the players' improvement of their workout performance with a dumbbell.

On the other hand, [12] simulate the football environment using MATLAB with the pre-determined scenario. Through the computer algorithm, efficient data collection of athletes on the fields during the competition process is achieved, providing more accurate data for tactical design and new ideas for athletes' training planning. [13] develop a cricket ball embedded with ESP8266 microcontroller and MPU6050 sensor capable of capturing the kinematics of the ball. The kinematics data are computed to extract performance data such as the ball's speed, spin, and impact location. Players can receive real-time feedback on their performance during practice and games.

Apart from player performance, psychological health is also essential in sports, especially during training sessions. [14] and [15] reviewed related studies on psychological parameters monitoring systems, which can also accurately detect athlete performance. Differently, [16] exploring the awareness level of IoT adaptation in sports among athletes is also important. They highlight the increasing perception and acceptance among athletes towards health and safety in IoT-integrated indoor systems.

The literature has summarized the adoption of IoT in sports, which focuses on improving the athlete's performance, monitoring the health status, and facilitating coaching sessions. However, none of them studied the potential use of IoT during competition. One element is in boule sport, where the distance of the thrown ball is still manually measured, which is time-consuming and prone to human error. For example, the current tool to measure the distance in Petanque sport still uses a ruler and measuring tape, which seems outdated and can lead to unfair judgment if misuse of the measurement tool. For this reason, this study attempts to test the application of sensor technology for distance measurement in Petanque sport and possibly gives an initial idea for future work applications. Measuring distance was studied in different domains, such as vehicle distance detection [17], social distance monitoring [18] and waste management [19] which promises to be used in boule sports such as Petanque.

## 3. METHODOLOGY

The Rapid Application Development (RAD) model is selected for this project methodology because it provides rapid prototyping and iterative delivery. Therefore, the RAD methodology will reduce the time for the development process. There are four main phases in RAD methodology: requirements planning, user design, construction, and cutover. However, this project only considers the first three phases since the final phase is only when the prototype is fully ready.

*Phase I: Requirements Planning*

This phase is essential to carrying out the project without any significant issues or problems. It is necessary to properly plan the entire project and analyze the problem statements, scopes, and objectives to fully understand requirements before proceeding with the project. The software and hardware requirements are also gathered to accomplish the project in this phase. For this project, the software and hardware requirements are stated below.

These are the two main software that are required for this project:

   a) Arduino IDE – A software for writing and uploading code into the microcontroller board.
   b) Blynk Application – IoT platform that interconnects microcontroller to the Internet.

The hardware required for this project is listed below:

a) Node MCU – A development kit microcontroller to execute an uploaded program for the project.
b) HC-SR04 Ultrasonic Sensor – An ultrasonic sensor is developed to detect an object using sound waves. The ultrasonic ranging module HC-SR04, a low-cost ultrasonic sensor, is the most commonly used. It provides a non-contact measurement function and can measure up a range from two cm to 400 cm with a ranging accuracy that reaches three mm. The ultrasonic sensor generates sound waves using a specific frequency above the human audible range, roughly 40 kHz.
c) Accessories such as jumper wire – connect the sensor with the microcontroller on the breadboard, breadboard – hold the electrical components for connection, power bank – power source for the device and micro-USB cable – connect the microcontroller with power source and for connection of microcontroller with a computer.

*Phase II: User Design*

In this phase, the jack and mobile application prototype is designed. Figure 1 shows the prototype of the jack, consisting of an ultrasonic sensor and nodeMCU board. When the player throws the boule, the ultrasonic sensor emits a sound pulse beyond the human hearing range (ultrasonic). This pulse travels toward the target (boule), and the sensor calculates the time it takes for the echo to bounce back, determining the object's position from the jack. The nodeMCU will then process the sensed data and transfer it to the mobile application via a Wi-Fi connection.

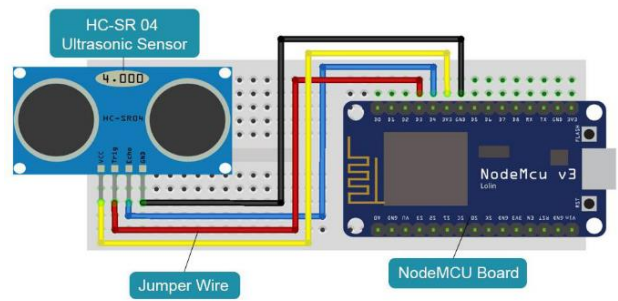

Figure 1. Prototype of a jack

The flowchart of the prototype system is depicted in Figure 2. The system is implemented based on the official rule of Petanque sport for the player to turns and score points. It considers two players, with each of them being given three boules to throw alternately. The result comparison between the players determines the score result.

*Phase III: Construction*

The construction phase works along with the user design phase, where the configuration for the microcontroller and connection with the sensor is constructed according to the circuit design that has been developed in the user design phase. The development continues to develop the mobile application, which will follow with testing. Figure 3 shows the testing phase settings for the jack, boule, and mobile application.

Figure 2. Flow chart of the prototype system

Figure 3. Experiment setting

## 4. TESTING AND RESULTS

A few types of testing are functionality, accuracy & reliability, and usability. Each of the tests is elaborated on in the following.

*i. Functionality Test*

The functionality test ensures the prototype is integrated with the NodeMCU and that the ultrasonic sensor works properly. The testing began with the testing of the NodeMCU to obtain distance measurement data from the sensor, send the data to the mobile application through a wireless connection, and display the data in the smartphone. Table 1 shows an example of test case TC002, which tests the connection of elements of the prototype jack.

Table 1. Example test case

| Test Case ID | TC 002 |
|---|---|
| Test Functionality | Test the connectivity of NodeMCU with the HC-SR04 ultrasonic sensor. |
| Precondition | The physical connection between NodeMCU and the sensor is configured. |
| Execution Steps | 1. Ensure the program script is uploaded into the NodeMCU board using Arduino IDE software.<br>2. Open the serial monitor on Arduino IDE.<br>3. View the sensor reading. |
| Expected Result | The ultrasonic sensor produces the measurement of the distance displayed in the Arduino IDE serial monitor. |
| Error Message | None |
| Result | Pass |

*ii. Accuracy and Reliability Test*

The accuracy and reliability test was conducted in a real environment mimicking the Petanque game. To begin the distance measurement accuracy test, the prototype and the environment should be simulated as in a real scenario of the Petanque game, in which the measuring object is the boule and the jack is the sensor. Figure 4 shows the initial setup for the testing, with the manual measurement using the ruler while the automated measurement is via the prototype jack, and it will display the measurement output in the smartphone.

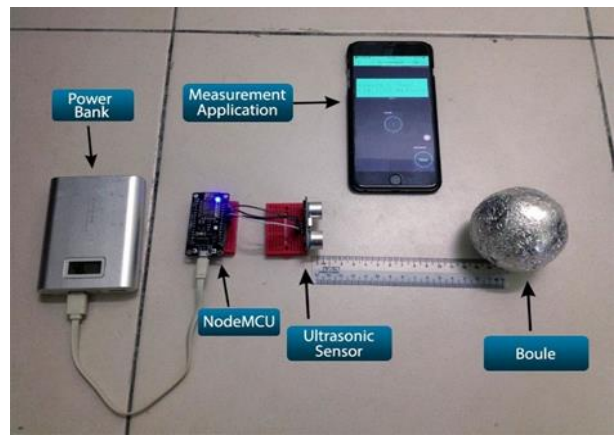

Figure 4. Initial setup for accuracy testing

Distance measurement accuracy testing is carried out by comparing the manual and automated measurements. The testing starts with a boule placement at three centimeters from the jack and continues with five and ten centimeters. The time taken for manual measurement is around 10 to 15 seconds, while reduced with the prototype jack, measurement results appear in the mobile application between 3 to 5 seconds. Two testing settings are conducted: indoor and outdoor.

Details of each test case are as follows: -

Test Case 1 - Indoor:

In this test case, the prototype is set in the indoor environment with the following criteria:
a) Secure in the indoor environment.
b) Not interrupted by wind and noise.
c) Manual distance measurement is fixed and measured by a ruler during the testing.
d) The automated measurement is an output from the prototype jack.

Test Case 2 - Outdoor:

In this test case, the prototype is set in the outdoor environment with the following criteria:
a) Expose to the outdoor environment.
b) Interrupted by wind and noise.
c) Manual distance measurement is fixed and measured by a ruler during the testing.
d) The automated measurement is an output from the prototype jack.

The result of the data accuracy test is presented in Table 2. The distance range covers between 3.0 cm to 15.0 cm, mimicking the play area of the petanque game. The standard deviation value shows the consistency of reading from the prototype jack compared to the manual measurement. The mean value for the standard deviation indoors is 0.03 cm, while for outdoors, it is much larger, 0.05 cm. The open space impacts this situation outdoors due to environmental factors such as air temperature and wind movement.

Table 2. Testing result of distance (cm) for indoor and outdoor

| Actual Distance (Manual Method) | Indoor - Prototype | | | | Outdoor - Prototype | | | |
|---|---|---|---|---|---|---|---|---|
| | Test 1 | Test 2 | Test 3 | Std. Deviation | Test 1 | Test 2 | Test 3 | Std. Deviation |
| **3.0** | 3.01 | 3.00 | 3.00 | 0.01 | 3.01 | 3.03 | 3.03 | 0.03 |
| **5.0** | 5.00 | 5.00 | 5.04 | 0.04 | 5.05 | 5.04 | 5.02 | 0.05 |
| **10.0** | 10.04 | 10.01 | 10.02 | 0.04 | 10.04 | 10.05 | 10.04 | 0.05 |
| **15.0** | 15.03 | 15.03 | 15.01 | 0.03 | 15.06 | 15.03 | 14.99 | 0.06 |
| | **Mean of Std. Deviation** | | | 0.03 | | | | 0.05 |

*iii. Usability Testing*

The usability testing evaluates the use of the prototype system on the actual game of Petanque sport. The test required two players to participate in this testing, who will play the Petanque sport based on the official rules. Each of the players will be given three boules; the testing begins by making the players start the play of the Petanque game. Figures 5(a) and 5(b) show the player's turn, which the mobile application can determine. In Figure 5(c), the winner of round 1 is notified.

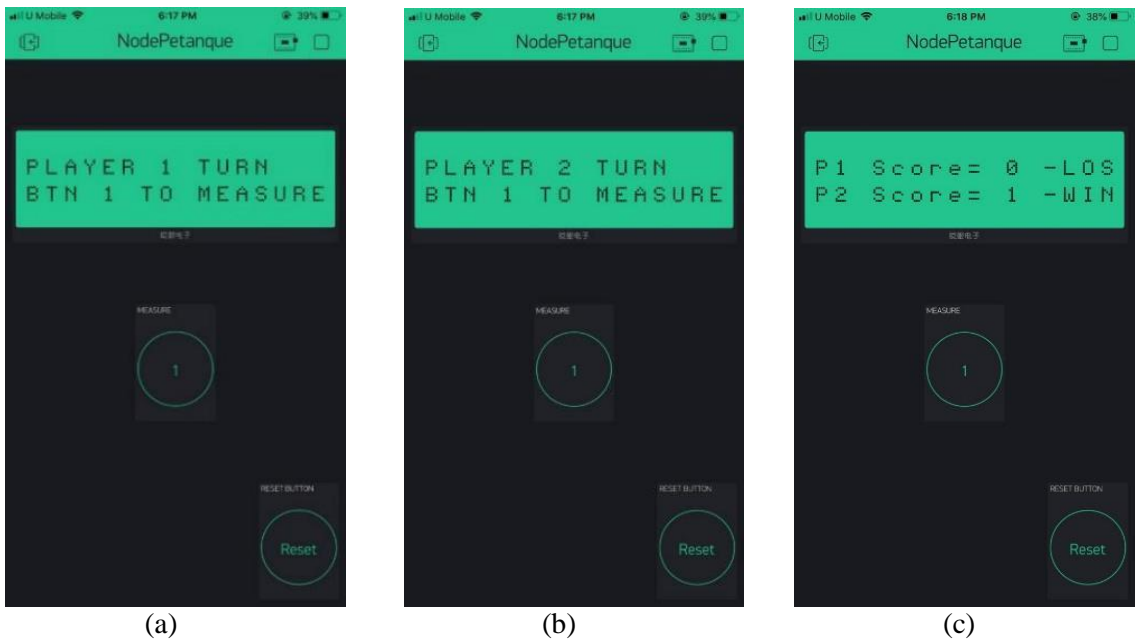

|  (a)  |  (b)  |  (c)  |

Figure 5. Mobile app interfaces.

The feedback from the usability testing by the player are:

a) Incomplete function – adding a specific button for each boule to measure the new measurement if the recorded measurement is changed due to shooting strategy plays.
b) Improve the result button by considering the winner for the overall game.

## 5. CONCLUSION

The prototype distance measurement device aims to ensure that this prototype device's use can help the fair judgment of distance measurement in Petanque sport. This project also aims to ease the method of distance measurement of the boule and jack in Petanque by simply pointing the ultrasonic sensor to the boule from the jack. The technological method has shown great results over the years compared to the manual method. Other than that, the distance measurement will be instantly calculated and displayed on the smartphone through a mobile application, which can save time when comparing the result data of the game.

This research has shown the possibility of IoT-based systems facilitating boule sport. However, this prototype is incomplete, in both mobile application and prototype jack. A suitable sensor is needed to improve distance reading in indoor and outdoor environments. Thus, it required further exploration.

### ACKNOWLEDGEMENTS

The authors would like to thank Information Security Forensics and Computer Networking Research Group, Center for Advanced Computing Technology (C-ACT), Fakulti Teknologi Maklumat dan Komunikasi (FTMK), Universiti Teknikal Malaysia Melaka.

## AUTHORS

**Wahidah Md Shah** holds her Bachelor of Information Technology from Universiti Utara Malaysia, Master of Computer Science from Universiti Teknologi Malaysia and PhD in Computer Science from Lancaster University, UK. She is currently a Senior Lecturer in the Department of Computer System and Communication at Universiti Teknikal Malaysia Melaka. She is a member of the Information Security, Digital Forensic, and Computer Networking research group. Her research interests include system and networking, wireless ad-hoc networking, cyber-physical systems (CPS) and IoT related technology. 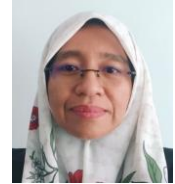

**Muhammad Azim** holds a Diploma in Networking from Politeknik Ungku Omar and a Bachelor's Degree in Computer Science (Networking) from Universiti Teknikal Malaysia Melaka. He has three years of experience as a small business's sole IT professional, honing his system management and troubleshooting skills. He plays a key role in overseeing the IT operations at Alila Bangsar Kuala Lumpur, where he has contributed to the hotel's success for nearly two years. 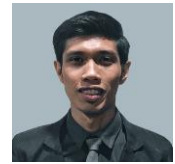

**Aslinda Hassan** received her PhD degree in Electrical Engineering, from Memorial University of Newfoundland, St. John's, NL, Canada in 2014. She received M.Sc. degree in Computer Science, from Universiti Teknologi Malaysia (UTM) and B.Sc. degree in Business Administration with honors, from University of Pittsburgh, Pittsburgh, PA, USA in 2001 and 1999, respectively. In 2004, she joined Universiti Teknikal Malaysia Melaka, where she is currently a Senior Lecturer at Faculty of Information and Communication Technology. Her research interests include in vehicular ad hoc network, vehicular communication, wireless ad-hoc network, wireless sensor network, wireless communication, ad hoc routing protocols, cyber-physical systems (CPS), Internet of Things (IoT), network performance modelling and analysis as well as network programming interfaces. 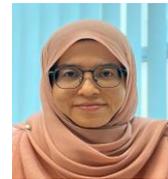

**Norharyati Harum** holds her Bachelor in Engineering (2003), MSc. in Engineering (2005) and PhD in Engineering (2012) from Keio University, Japan. She has experience working in R & D Department of Next Generation Mobile Communication at Panasonic Japan (2005-2009). She is currently a senior lecturer at Faculty of Information and Communication Technology, UTeM. Her interests in research area are Internet of Things, Wireless Sensor Network, Next Generation Mobile Communication and Signal Processing. She is an accomplished inventor, holding patents to radio access technology, and copyrights of products using IoT devices. 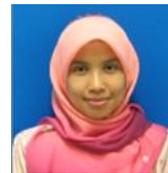

**Isredza Rahmi A Hamid** Isredza Rahmi A Hamid holds her Bachelor of Information Technology from Universiti Utara Malaysia, Master Science of Information Technology from University Teknologi MARA and PhD in Information Technology from Deakin University, Australia. She is currently a Senior Lecturer in Information Security and Web Technology Department at Universiti Tun Hussein Onn Malaysia. She is a Principle Researcher of the Information Security Interest Group (ISIG). Her research interest includes Information Security, Malware Detection, Clustering Algorithm, Data Mining and Soft Computing. 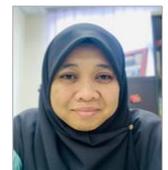